# Degradation mechanism of 0.15 μm AlGaN/GaN HEMTs: effects of hot electrons

Z. Gao*, F. Rampazzo, M. Meneghini, C. De Santi, F. Chiocchetta, D. Marcon, G. Meneghesso, E. Zanoni

*Abstract – The degradation mechanisms of AlGaN/GaN HEMTs adopting Fe and C co-doping, with high and low carbon doping concentration were investigated by means of hot-electron step stress and 24 hours' stress tests. Firstly, DC and EL characterization at room temperature are summarized, then the parametric evolution during hot-electron step stress tests at the semi-on state was compared, the assumption for the degradation mechanism is that hot-electrons activated the pre-existing traps in the buffer, attenuate the electric field in the gate drain access region and damaging the gate contact, the parametric evolution during constant stresses is discussed.*

## 1. Introduction

Development of GaN High Electron Mobility Transistors (HEMTs) for telecom, radar and space applications has been improving steadily [1]–[4]. The demand for industrial-level 0.15 μm HEMTs for low-noise and power amplifiers up to and beyond 50 GHz [5] has been enhanced by 5G and radar applications. In order to control short-channel effects, which directly determine leakage current and RF reliability [6], various approaches have been used, including back-barriers [7], C or C+Fe co-doping [8], N-polar GaN/AlGaN channels [9] and so on. Reliability is another key issue that has to be taken into consideration during the improvement of HEMT technologies. In the working condition of RF HEMTs, the devices will experience either high electric field (off-state), or high power density (on-state) or the combination of high current density and high electric-field (semi-on state).

The degradation mechanisms of the devices could be summarized as trap states generation [10], hot-electron generation [11], gate metal instabilities [12], passivation breakdown [13], inverse piezoelectric effects [14] and so on. For 0.15 μm gate length devices, the effects of electric field have proven to be important in reliability [8]. However, there is still not enough study on the combined effects of trapping centres, high electric field and hot electrons.

This work is a follow up study of [8], comparing 0.15 μm gate length GaN HEMTs with different epi design. The devices were fabricated within the same batch, using the same industrial level process, with optimized gate metallization and passivation, on different epi layers. The effects of hot electrons on the reliability of devices are studied using electroluminescence (EL), and comparing the parametric degradation of devices during hot-electron stress.

## 2. Experimental details

Three groups of devices were fabricated on AlGaN/GaN heterostructures grown on SiC wafer, with gate length of 0.15 μm and gate width of 100 μm, processed within the same batch using a standard RF GaN HEMT process, on three types of epitaxial layers from different manufactures. All the samples had similar nominal peak Fe concentration i.e. $2 \times 10^{18}$ $cm^{-3}$, but different carbon doping profile, one sample is non-intentionally doped, the other two have different carbon peak concentrations: $2 \times 10^{16}$ $cm^{-3}$ and $8 \times 10^{16}$ $cm^{-3}$, hereafter identified as "Fe Reference", "Fe + Low C" and "Fe + High C". Carbon co-doping is supposed to be helpful in reducing short-channel effects.

HEMTs were submitted to on-wafer step-stress and constant voltage 24h stress tests. During step stress (in semi-on state, where the EL intensity is the strongest), drain voltage was increased from 0 V up to devices catastrophic breakdown in 5 V step, two minutes long each. During each stress step, drain and gate currents as well as EL intensity were monitored. After each stress step, the device was kept unbiased for 5 minutes, afterwards, a standard DC characterization and an EL image at the bias condition where there is the EL Bell peak ($V_G$ = -1.5 V, $V_D$ = 5 V) was taken, as shown in Fig. 1 (a).

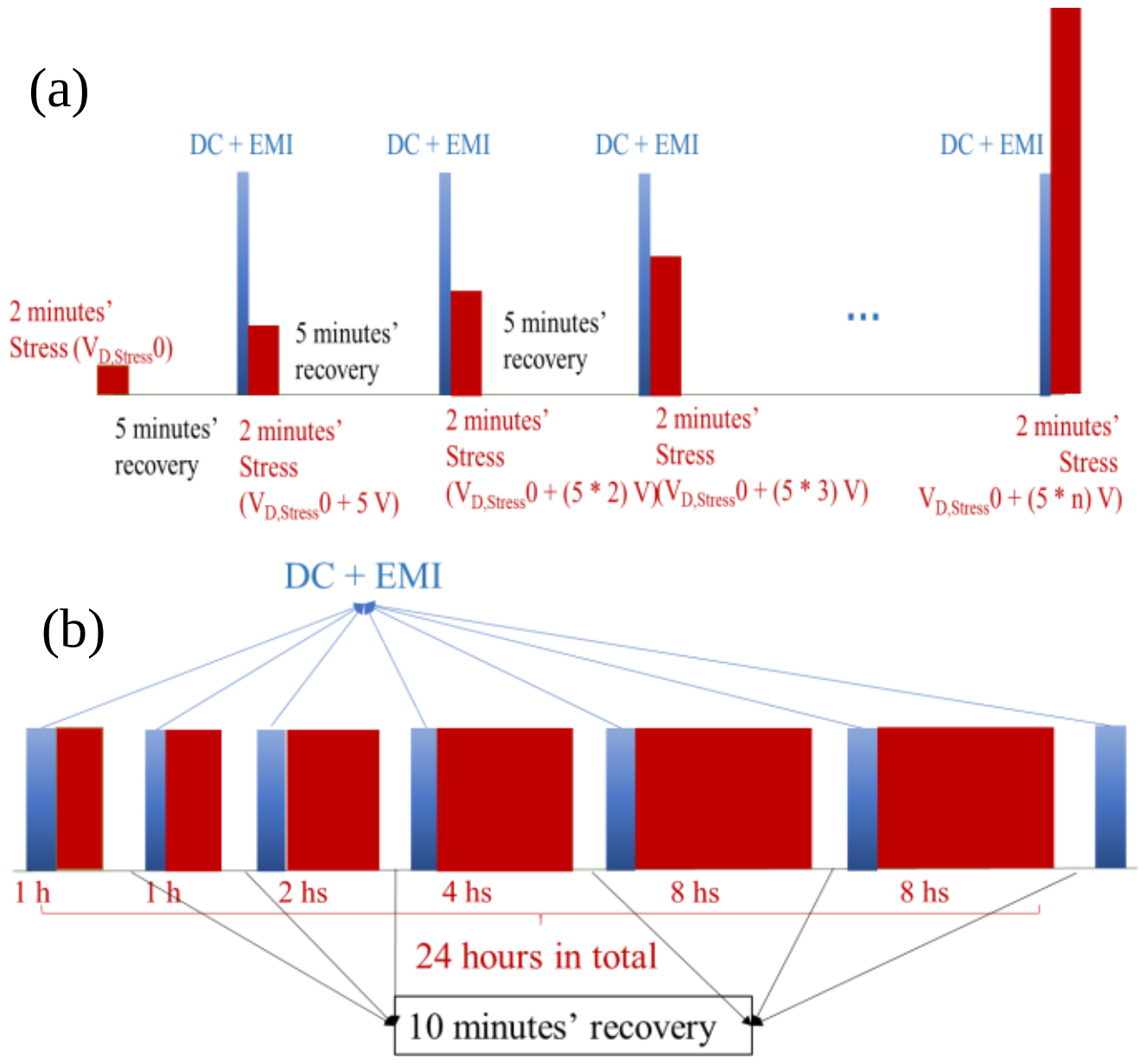

Fig. 1 Flow chart of (a) the step stress and (b) 24 hours' stress

Afterwards, 24 hours hot-electron stress were done on fresh devices, at gate voltage $V_{G,Stress}$ = -2 V, and drain voltage ($V_{D,Stress}$) at 10 V, 15 V, 20 V and 25 V. During stress, drain and gate currents were monitored, and a standard DC characterization were done after 1 hour, 2 hours, 4 hours and 8 hours, 16 hours and 24 hours' stress, after being kept unbiased for 10 minutes, as shown in Fig. 1(b). At least two devices were tested at each test condition.

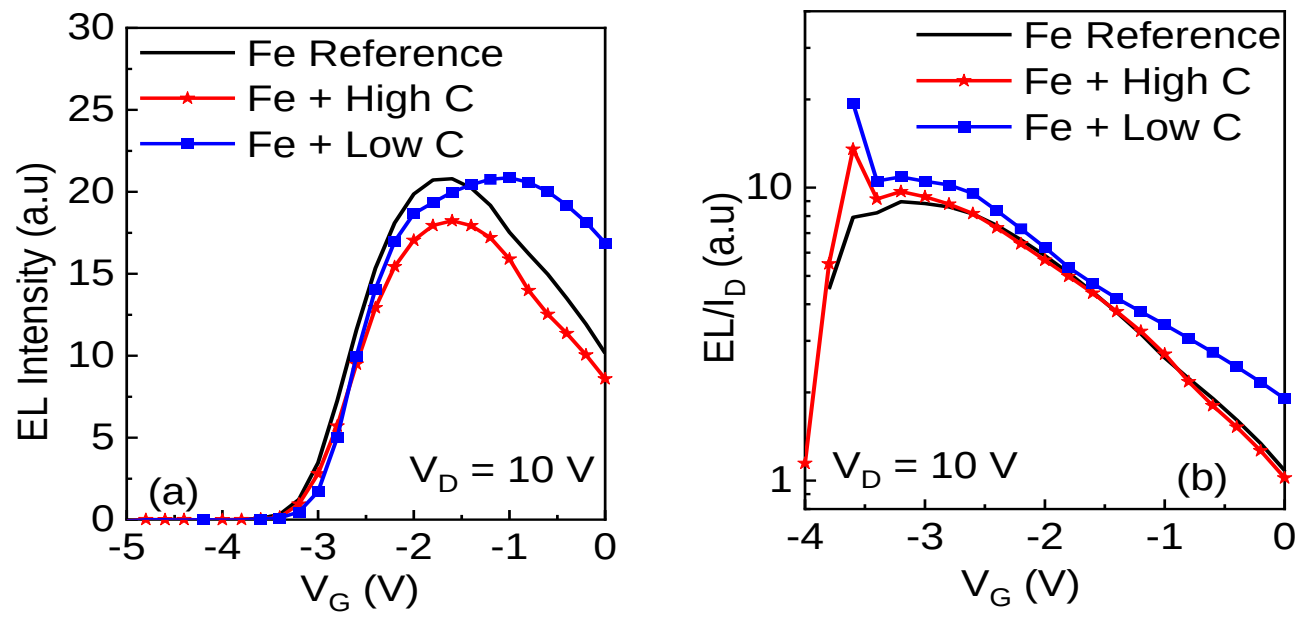

Fig. 2 EL intensity and EL/$I_D$ comparison of the three types of devices

## 3. Results and discussions

### *3.1 Preliminary results*

DC characteristics showed that the "Fe Reference" devices have the smallest leakage current; and the "Fe + High C" device has reduced Drain Induced Barrier Lowering (DIBL) effects due to the better charge confinement due to the high carbon doping concentration, as shown in [8]. The EL intensity follows a bell shape versus $V_{GS}$ biasing for all devices, as shown in Fig. 2. The bell peak lies close to -2 V, and the EL/$I_D$ ratio in on-state ($V_G$ > -1.5 V) indicating that the "Fe + Low C" samples have the highest electric field in the channel among the three types of devices, noticing a very small EL difference between the "Fe Reference" and the "Fe+ High C" devices.

### *3.2. hot-electron step stress*

Hot electron step stresses were done on the three types of wafers, with at least two devices from each wafer. The tests were done at gate stress voltage of -2 V, close to the EL peak and drain stress voltage from 0 V to breakdown. During stress, all the devices showed catastrophic failure at 45 V. All the devices showed less than 10 times parametric change in leakage current, and less than 10% maximum transconductance ($g_{m,max}$) decrease.

The maximum drain current ($I_{DS,max}$) and threshold voltage ($V_{TH}$) shift are shown in Fig. 3(a). $V_{TH}$ shift negatively with increasing $V_{D,Stress}$, -0.11 V at $V_{D,Stress}$ = 20 V for "Fe-High C" wafer, -0.06 V at $V_{D,Stress}$ = 25 V for "Fe-Low C" wafer and -0.12 V at $V_{D,Stress}$ = 30 V for Fe referemce wafer. Afterwards, $V_{TH}$ shifted to the opposite direction with increasing $V_{D,Stress}$, and $I_{DS,max}$ starts to decrease. The Fe Reference wafer showed a $I_{DS,max}$ decrease less than 10%, while the other two wafers showed over 10% $I_{DS,max}$ decrease and 10% $R_{ON}$ increase at drain stress voltage of 35 V and 40 V. At $V_{D,Stress}$ = 45 V, positive $V_{TH}$ shift up to +0.5 V and +0.3 V are observed for the "Fe-High C" and "Fe-Low C" wafer, respectively.

Drain current monitored during stress are shown in Fig. 3 (b), there is continuous increase during stress from 0 V to a certain voltage, 20 V for the "Fe-High C" wafer, 25 V for the "Fe-Low C" wafer and 30 V for the "Fe reference" wafer.

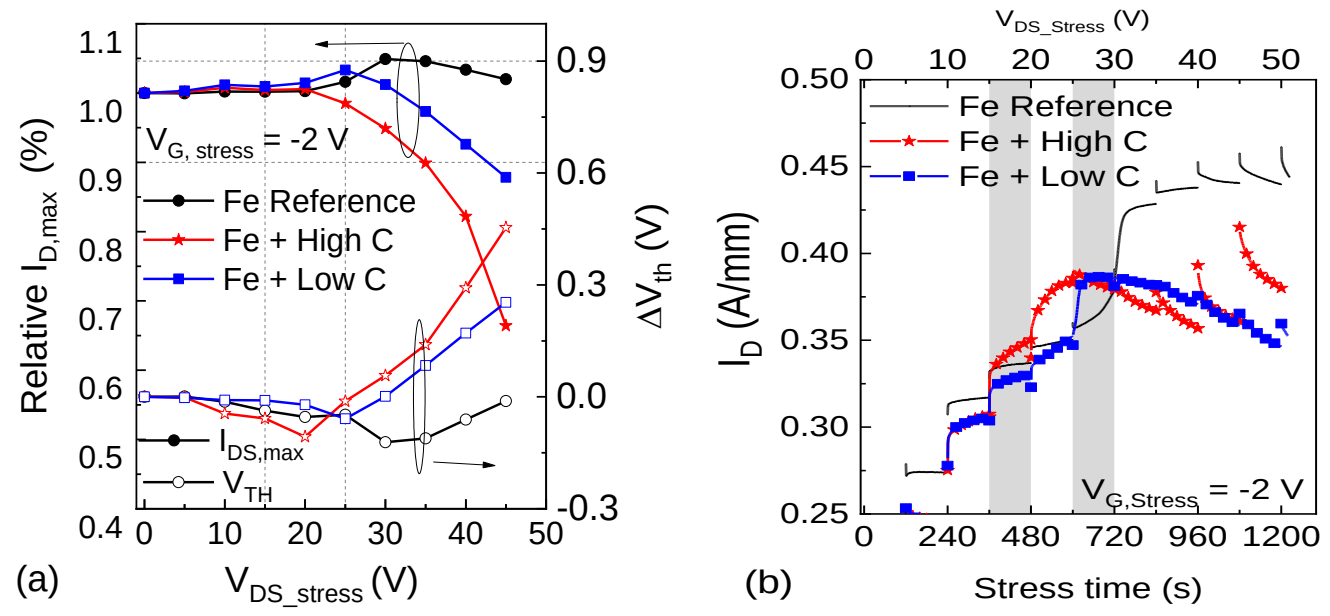

Fig. 3. (a) normalized $V_{th}$ and $I_{DS,max}$, and (b) Drain current during stress during hot electron step stress at $V_G$ = -2 V.

EL intensity, taken at the EL Bell peak after each step during stress, decreases with increasing stress voltage, as shown in Fig. 4(a). In order to identify the cause of the EL intensity decrease, relative EL/$I_D$ at this bias point ($V_G$ = -1.8 V, $V_D$ = 5 V) were shown in Fig. 4 (b). EL/$I_D$ decreased up to certain voltage, 20 V for the "Fe + High C" wafer, 25 V for the "Fe + Low C" wafer, and 30 V for the "Fe reference" wafer, then it shifted in the opposite direction up to breakdown.

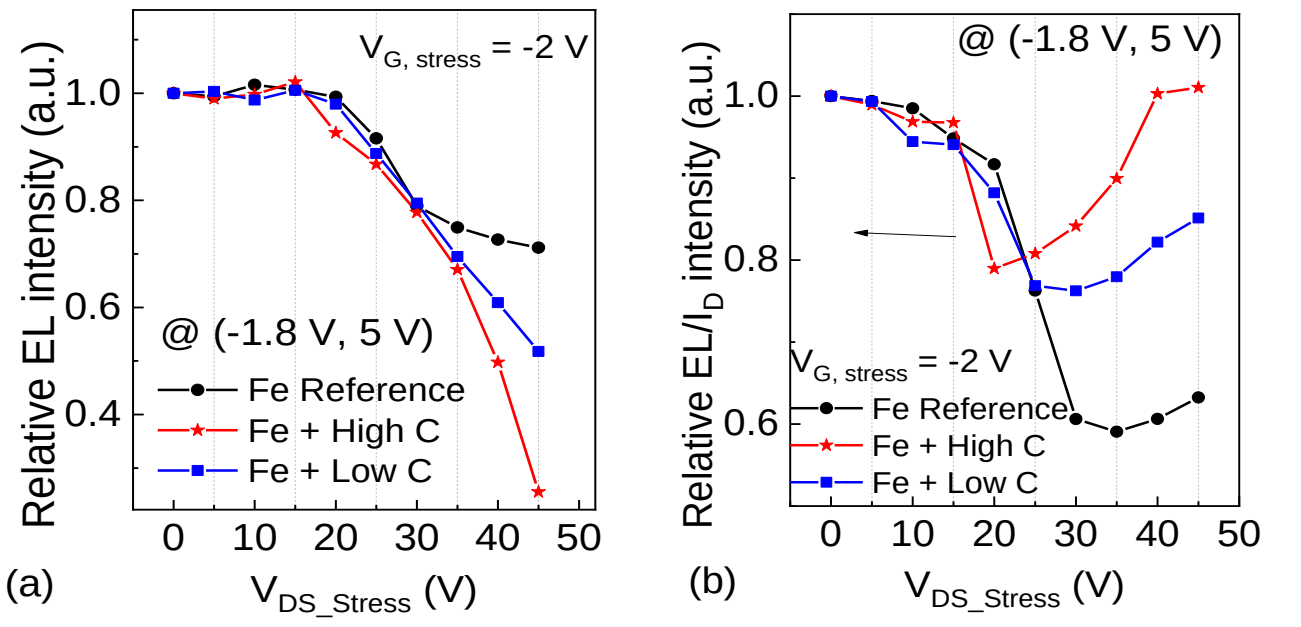

Fig. 4 (a) El intensity and (b) EL/$I_D$ at ($V_G$ = -1.8 V, $V_D$ = 5 V) during hot electron step stress at $V_G$ = -2 V.

The EL image taken at off-state (-5 V, 10 V) after stress at (-2 V, 25 V) are shown, there are four leaky points on the "Fe + High C" wafer, and one leaky point on the "Fe + Low C" wafer, in the gate-drain access region, close to the gate contact, which was not observed before stress. There was almost no parametric degradation in the Fe-reference wafer up to 45 V. As the three wafers shared the same processing steps and were produced within the same batch, this suggests that there are no process-related degradation mechanisms, and would exclude, in particular, metal-semiconductor interdiffusion or surface electrochemical oxidation mechanisms.

### *3.2. hot-electron 24 hours stress*

24-hours stress tests were carried out, at bias points at semi-on states, with $V_G$ = -2 V and $V_D$ from 10 V to 25 V with 5 V step. The DC characteristics evolution during stress at (-2 V, 15 V), (-2 V, 20 V) and (-2 V, 25 V) is discussed in the following sections.

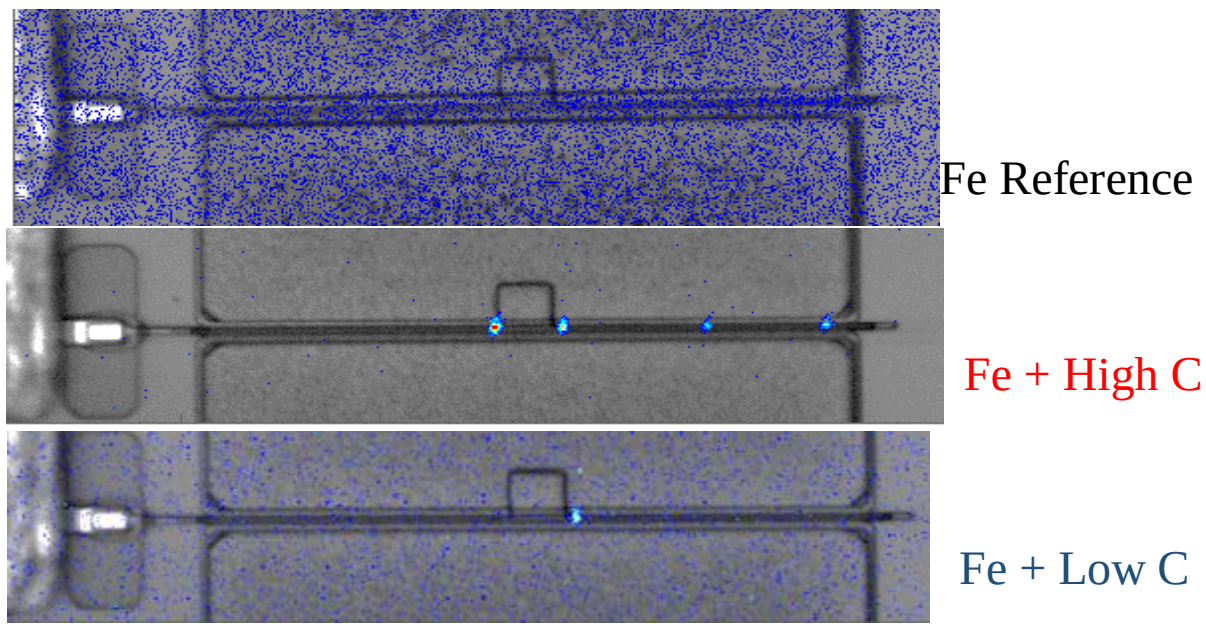

Fig. 5 El image at ($V_G$ = -5 V, $V_D$ = 10 V) after stress step at ($V_G$ = -2 V, $V_D$ = 25 V) of the three wafers.

*3.2.1 hot-electron long term stress at (-2 V, 15 V)*

DC characteristics change during stress at (-2 V, 15 V), $\Delta I_{DS,max}$ and $\Delta V_{TH}$ evolution as a function of stress time are shown in Fig. 6.

All the wafers showed bell shape in $I_{DS,max}$ evolution with stress time, in accordance with the $V_{TH}$ shift during stress. The $V_{TH}$ shifts after one hour's stress are about -0.14 V, -0.12 V and -0.11 V for "Fe Reference", "Fe + Low C" and "Fe + High C" wafers respectively. Afterwards, the $V_{TH}$ showed right shift in the "Fe + High C" and "Fe + Low C" wafers, the total $V_{TH}$ shift after 24 hours' are +0.14 V and + 0.04 V for the "Fe + Low C" and "Fe + High C" wafers, respectively. For the "Fe Reference" wafer, the negative $V_{TH}$ shift continued after 2 hours' stress, up to -0.19 V, then it shifted in the opposite direction, the final $V_{th}$ shift being about -0.04 V.

In order to better understand the current change during stress, the normalized drain current transient are recorded during stress, as shown in Fig. 6 (b), there is current increase at the beginning, one hour for the "Fe + Low C" and "Fe + High C" wafers, and two hours for the "Fe reference" wafer. Afterwards, the current decreased with stress proceeds.

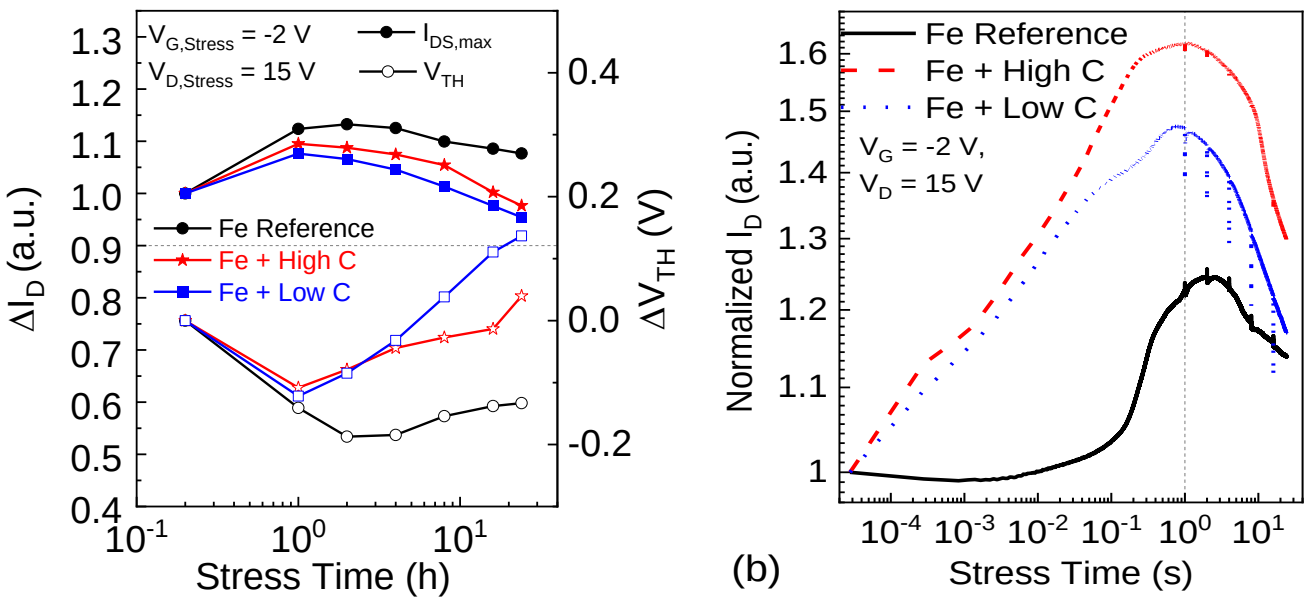

Fig. 6 (a) normalized $I_{DS,max}$, and $V_{TH}$ evolution and (b) normalized drain current transient during 24 hours' hot-electron stress at $V_G$ = -2 V, $V_D$ = 15 V.

The variation of Schottky barrier height (SBH) and gate leakage current ($I_{G,leak}$) at $V_{GS}$ = -7 V, $V_{DS}$ = 10 V during stress is shown in Fig. 7. The leakage current increases when the SBH decrease, and the leakage current decrease when the SBH turns to increases.

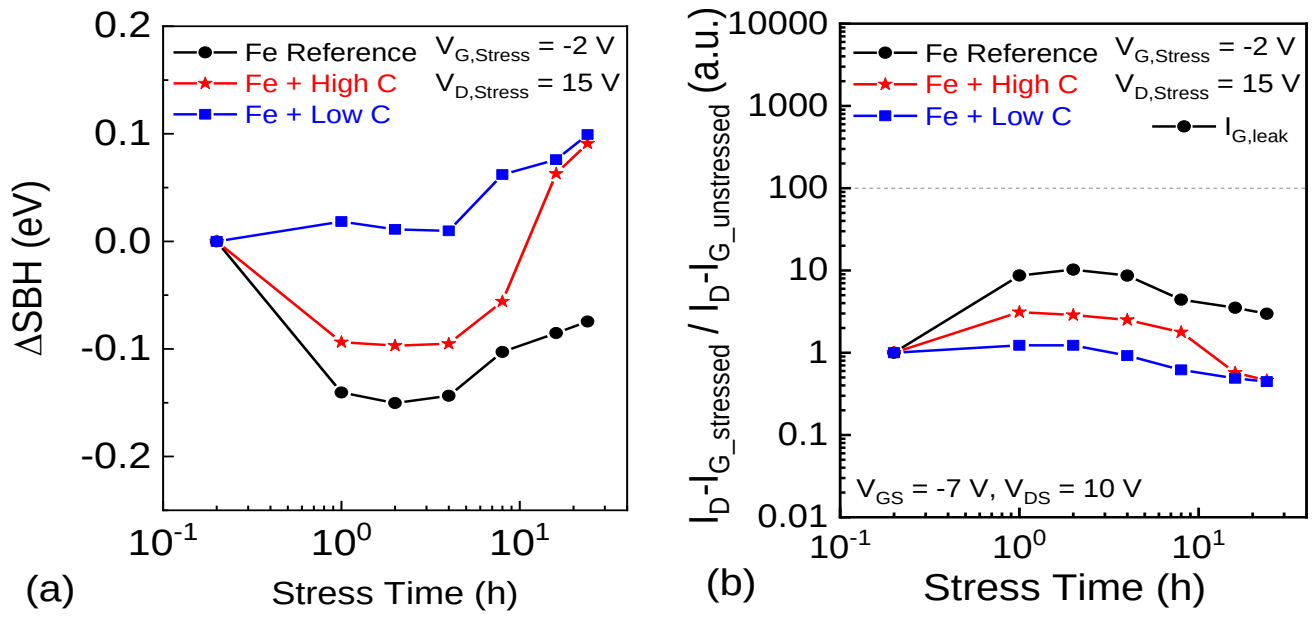

Fig. 7 (a) normalized SBH evolution and (b) normalized drain leakage current during 24 hours' hot-electron stress at $V_G$ = -2 V, $V_D$ = 15 V.

The transfer $g_m$-$V_G$ curves of the devices before and after 24 hours' stress are shown in Fig. 8. No significant $g_{m,max}$ decrease was observed after the 24 hours' stress in all the wafers, however, there is about 5%, 8% and 9% drop in transconductance at high $V_{GS}$ for "Fe Reference", "Fe + Low C" and "Fe + High C" wafers, respectively. In order to observe the effects of hot-electron stress, the same device from each wafer was biased at the same point for another 24 hours, results showed that there is 2%, 6% and 7% degradation in $g_{m,max}$ for the "Fe Reference", "Fe + Low C" and "Fe + High C" wafers, respectively; at the same time, the $g_m$ degradation at high $V_{GS}$ turn into saturation.

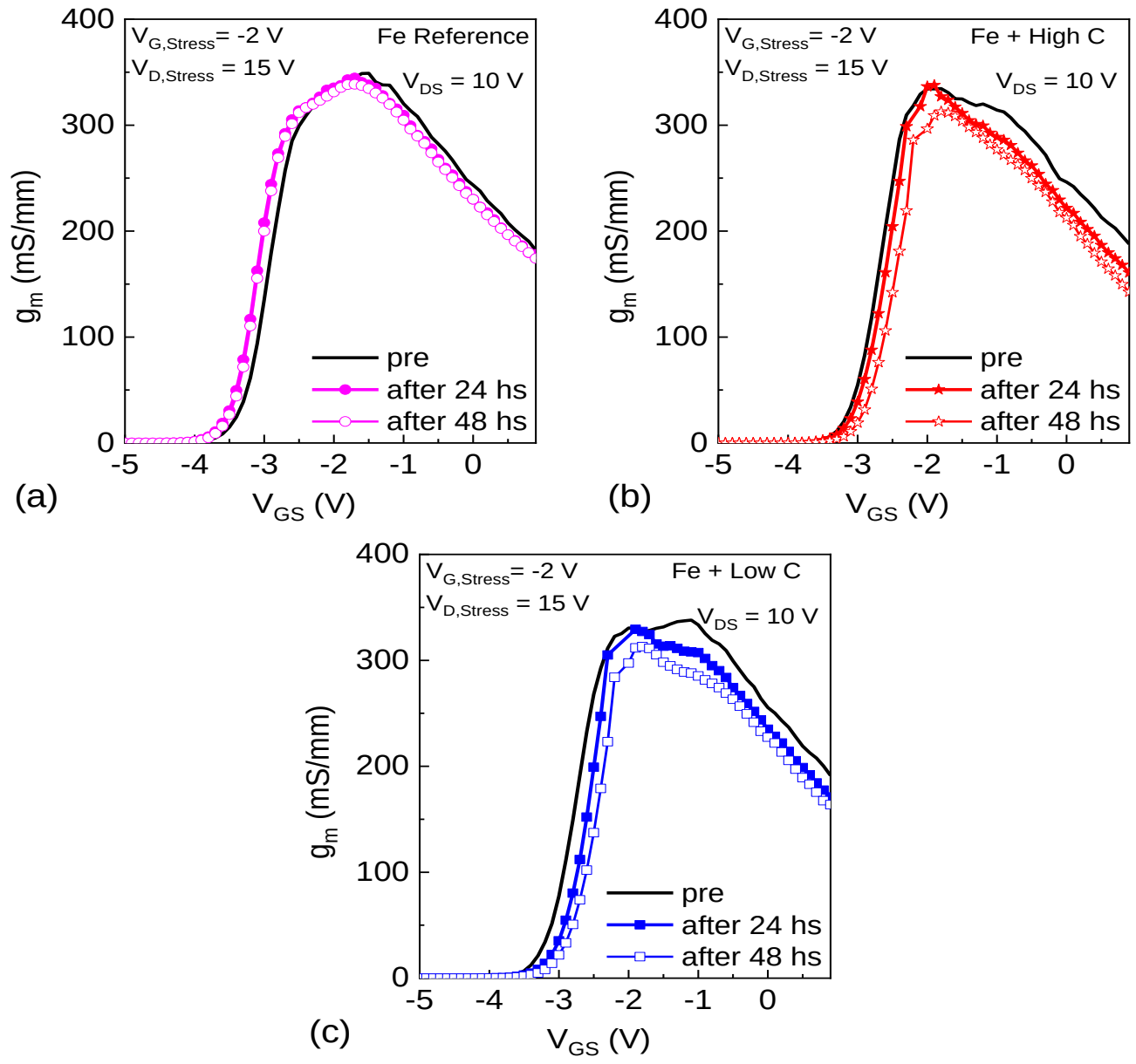

Fig. 8 Transfer $g_m$-$V_G$ curves of (a) "Fe reference" and (b) "Fe + High C" and (c) "Fe + Low C"wafers before and after 24 hours' stress at (-2 V, 15 V).

*3.2.2 hot-electron 24 hours stress at (-2 V, 20 V)*

DC characteristics change during stress at (-2 V, 20 V), $\Delta I_{DS,max}$, $\Delta R_{ON}$ and $\Delta V_{TH}$ evolution as a function of stress time are shown in Fig. 9Fig. **10**(a), and normalized drain current transien recorded during stress is shown in Fig. 9(b). Similar to the stress at (-2 V, 15 V), the drain current increased at the beginning of the stress, however, the increase trend ended within one hour, for all three types of wafers, then decreased at different rate after that.

The DC characteristics showed bell shape variation in the wafers. After 24 hours, $I_{DS,max}$ decreased by 6% and 4% and $V_{TH}$ shifted by +0.11 V and +0.14 V, in the “Fe + High C” and “Fe + Low C” wafer, respectively. The $V_{TH}$ shifted negatively by -0.13 V after the first hour’s stress in the “Fe Reference” wafer, then shifted slowly to the opposite direction, the final $V_{TH}$ shift is about -0.04 V after 24 hours’ stress.

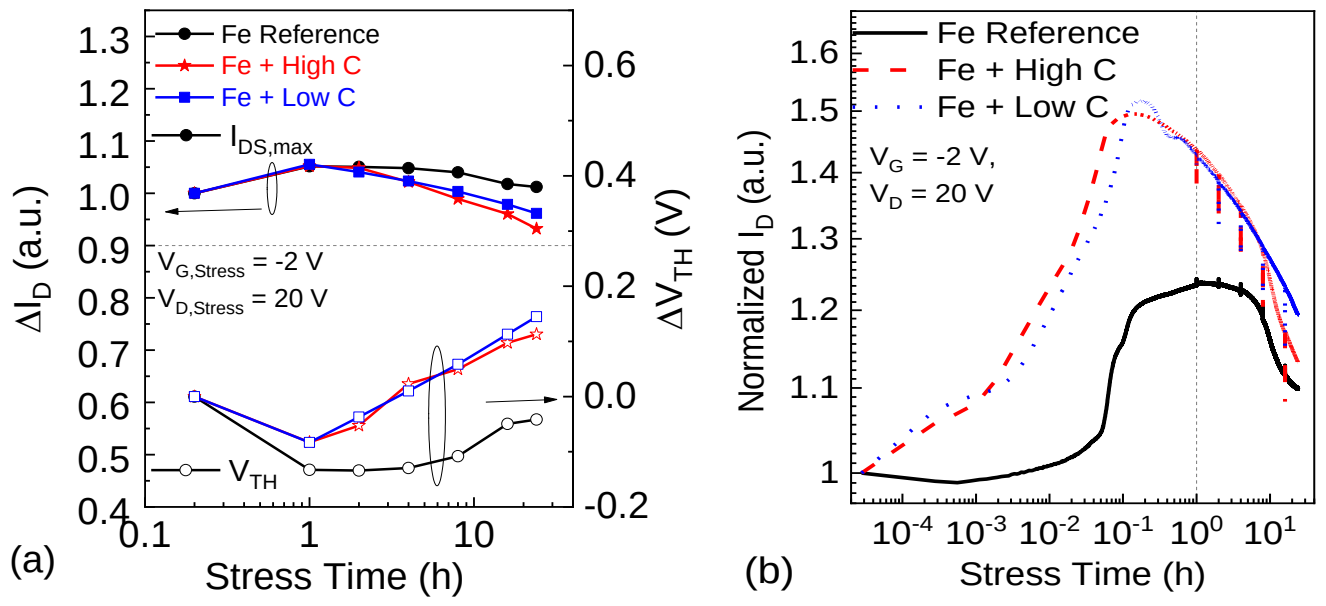

Fig. 9 (a) normalized $I_{DS,max}$ and $V_{TH}$ evolution and (b) normalized drain current transient during 24 hours’ hot-electron stress at $V_G$ = -2 V, $V_D$ = 20 V.

*3.2.3 hot-electron 24 hours stress at (-2 V, 25 V)*

DC characteristics change during stress at (-2 V, 25 V), $\Delta I_{DS,max}$, $\Delta R_{ON}$ and $\Delta V_{TH}$ evolution as a function of stress time are shown in Fig. 10. Monotonic variation was observed during the stress in the “Fe + High C” and “Fe + Low C” wafers. $I_{DS,max}$ decreased 10% and 8% and $V_{TH}$ shifted by +0.29 V and +0.38 V after 24 hours, in the “Fe + High C” and “Fe + Low C” wafer, respectively. The $V_{TH}$ shifted negatively by -0.12 V after the first hour’s stress in the “Fe Reference” wafer, then it shifted in the opposite direction slowly, the final $V_{TH}$ shift is about -0.06 V after 24 hours’ stress.

The normalized drain current transient is recorded during stress, as shown in Fig. 10(b). For the “Fe + Low C” and “Fe + High C” wafers, current increase occurred once stress started, and reached the maximum within 100 s, afterwards, there is drastic drop in drain current. For the “Fe Reference” wafer, there is gradual increase in drain current during stress, and it reached the peak within two hours, afterwards, there is gradual current decrease.

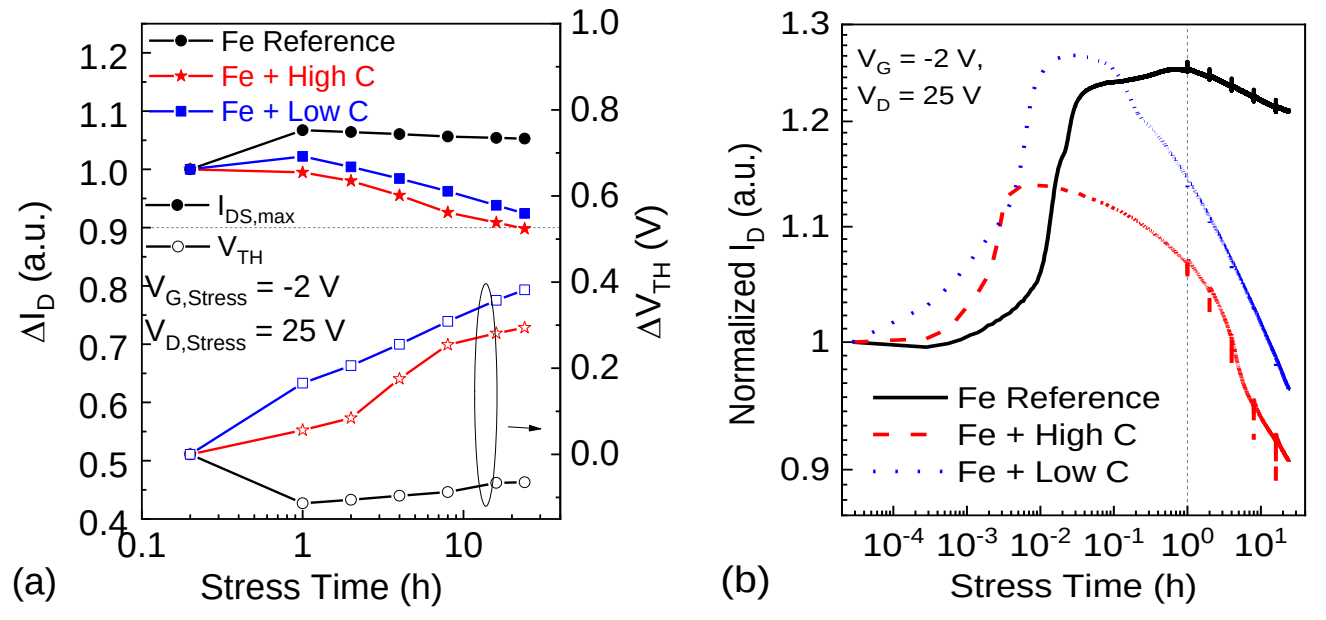

Fig. 10 (a) normalized $I_{DS,max}$ and $V_{TH}$ evolution and (b) normalized drain current transient during 24 hours’ hot-electron stress at $V_G$ = -2 V, $V_D$ = 25 V.

The variation of SBH and gate leakage current showed consistent trend during stress, similar to that observed in the device after stress at (-2 V, 15 V) and (-2 V, 20 V).

The transconductance curves before and after the 24 hours stress at (-2 V, 25 V) are shown in Fig. 11. There is 8% decrease of $g_m$ value at high $V_{GS}$ in the “Fe Reference” wafer. At the same time, $g_{m,max}$ decrease can be observed, the decrease percentage is about 3%, 10% and 6% for “Fe Reference”, “Fe + High C” and “Fe + Low C” wafers, respectively.

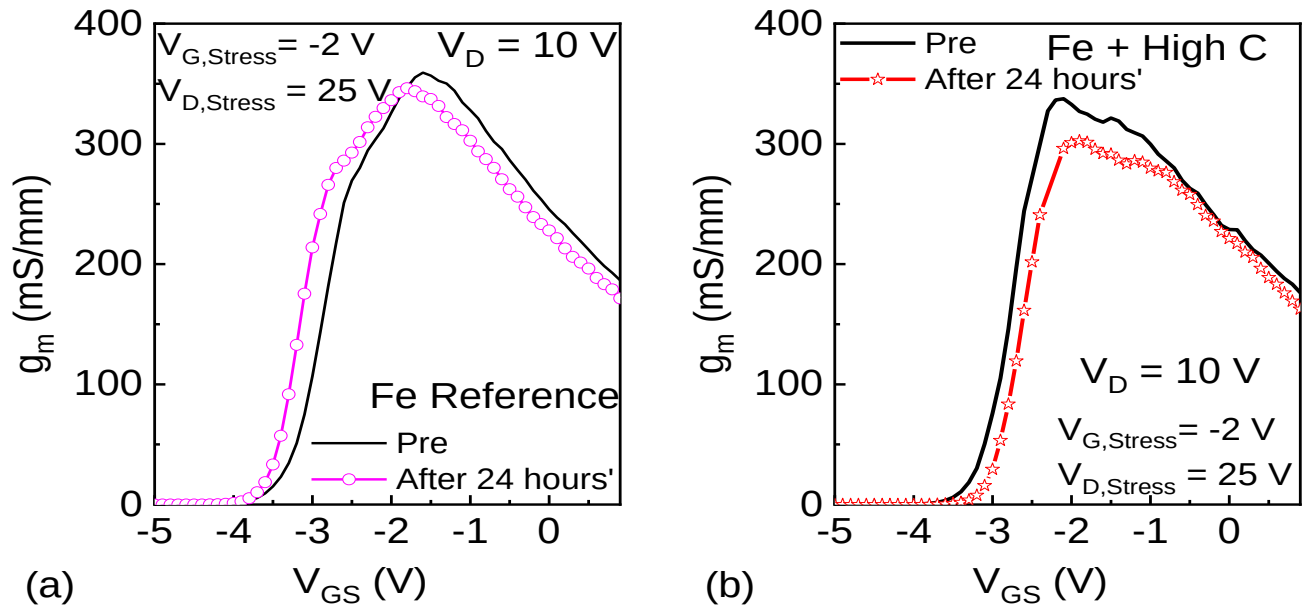

Fig. 11 Transfer $g_m$-$V_G$ curves of (a) “Fe reference” and (b) “Fe + High C” wafers before and after 24 hours’ stress at (-2 V, 25 V).

The dissipation power during stress at (-2 V, 15 V) and (-2 V, 20 V) is around to 3.5 W/mm and 5 W/mm, therefore the effects of self-heating can be ignored. During the stress at (-2 V, 25 V), the dissipation power is close to 8 W/mm, the channel temperature could reach 150°C, however, the parametric evolution of the devices follows similar trend to the devices stressed at other bias points, indicating that during this 24 hours’ stress, the device degradation is as a result of similar mechanism to the other two bias points, which is hot-electron effects, and thermal effects could be ignored in this case.

The results showed that, during stress, the three wafers presented drain current increase and negative $V_{TH}$ shift when stress started, then drain current decrease accompanied by positive $V_{TH}$ shift decrease after some time. However, the turning point and the final $V_{TH}$ shift value showed a clear difference among the three types of wafers, which is possibly caused by different doping profiles.

The different degradation modes of threshold voltage during stress can be explained as follows: The negative $V_{TH}$ shift and increase of drain current when stress started is possibly due to activation of hole traps by hot electrons in the GaN buffer close to the 2DEG [15], [16]; which lead to SBH decrease [17], gate leakage current increase and/or decrease of gate-edge electric field [9], which was observed as EL decrease, and the EL/$I_D$ decrease; and the positive $V_{TH}$ shift with increasing stress voltage or increasing stress time is due to the creation of traps in the channel or in the gate-drain access region [15]. And the transconductance decrease after 24 hours could be explained by that when the drain stress voltage is low, the damage is located at the drain contact side of the gate-to-drain access region (leading to transconductance collapse at high $V_{GS}$ values),

The hypothesis for the degradation is that the hot electrons activated during stress could remove the hydrogen from pre-existing defects, possibly $Fe_{Ga}$-$V_N$-H [18], [19], the dehydrogenation of the complex defect would lead to a negative $V_{TH}$ shift [20], and when $Fe_{Ga}$-$V_N$-H are depleted, the hot-electrons together with the electric field will gradually damage gate-drain access region close to the drain contact, leading to transconductance decrease, and when the drain stress voltage is high or the stress time is long enough, the damaged region extends to the gate contact at the gate-to-drain access region, leading to the collapse of $g_{m,max}$.

The difference during stress among the wafers with different carbon doping can be further explained by a parallel procedure: the effect of carbon related traps in the buffer and its interaction with hot-electrons. In the "Fe + High C" and "Fe + low" wafers, the carbon doping in the buffer are usually incorporated substitutional on either the N or Ga site, and $C_N$ as a deep acceptor and $C_{Ga}$ as a donor [21]–[24]. $C_N$ can be trapping in the GaN/AlN nucleation interface, band bending in GaN is therefore induced, leading to vacant $C_N$ levels, with the help of hot electron, the states can be filled, acting as back-gate, leading to positive $V_{TH}$ shift. The larger C doping concentration will make this procedure more likely and/or more rapidly to happen.

## 4. Conclusion

In conclusion, a comparison of parametric degradation among wafers with Fe and C co-doping varying C doping concentration is done, and the hypothesis accounting for the degradation is proposed. The degradation mechanisms of AlGaN/GaN HEMTs adopting Fe and C co-doping is discussed. The degradation is likely caused by dehydration effects of $Fe_{Ga}$-$V_N$-H complexes, and the effects carbon related traps. The assumption is supported by the results from hot-electron step stress tests and constant voltage stress tests.

## Acknowledgements

Support by EUGANIC project under the EDA Contract B 1447 IAP1 GP, by the EC Horizon 2020 ECSEL project 5G_GaN_2, and by the ESA ESTEC project RELGAN is gratefully acknowledged.

## References

[1] U. K. Mishra, P. Parikh, W. Yi-Feng, and Y. F. Wu, "AlGaN/GaN HEMTs-an overview of device operation and applications," *Proc. IEEE*, vol. 90, no. 6, pp. 1022–1031, 2002.

[2] E. Monaco, G. Anzalone, G. Albasini, S. Erba, M. Bassi, and A. Mazzanti, "A 2-11 GHz 7-Bit High-Linearity Phase Rotator Based on Wideband Injection-Locking Multi-Phase Generation for High-Speed Serial Links in 28-nm CMOS FDSOI," *IEEE J. Solid-State Circuits*, vol. 52, no. 7, pp. 1739–1752, 2017.

[3] E. Mammei *et al.*, "A power-scalable 7-tap FIR equalizer with tunable active delay line for 10-to-25Gb/s multi-mode fiber EDC in 28nm LP-CMOS," *Dig. Tech. Pap. - IEEE Int. Solid-State Circuits Conf.*, vol. 57, pp. 142–143, 2014.

[4] K. Hadipour, A. Ghilioni, A. Mazzanti, M. Bassi, and F. Svelto, "A 40GHz to 67GHz bandwidth 23dB gain 5.8dB maximum NF mm-Wave LNA in 28nm CMOS," *Dig. Pap. - IEEE Radio Freq. Integr. Circuits Symp.*, vol. 2015-Novem, pp. 327–330, 2015.

[5] V. Di Giacomo-Brunel *et al.*, "Industrial 0.15-μm AlGaN/GaN on SiC Technology for Applications up to Ka Band," *EuMIC 2018 - 2018 13th Eur. Microw. Integr. Circuits Conf.*, pp. 1–4, 2018.

[6] O. Breitschädel *et al.*, "Short-channel effects in AlGAN/GaN HEMTs," *Mater. Sci. Eng. B*, vol. 82, no. 1–3, pp. 238–240, May 2001.

[7] M. Micovic *et al.*, "High frequency GaN HEMTs for RF MMIC applications," in *2016 IEEE International Electron Devices Meeting (IEDM)*, 2016, pp. 3.3.1-3.3.4.

[8] Z. Gao *et al.*, "Reliability comparison of AlGaN/GaN HEMTs with different carbon doping concentration," *Microelectron. Reliab.*, vol. 100–101, no. May, p. 113489, Sep. 2019.

[9] D. Bisi *et al.*, "Observation of Hot-Electron and Impact-Ionization in N-polar GaN MIS-HEMTs," *IEEE Electron Device Lett.*, vol. 3106, no. c, pp. 1–1, 2018.

[10] C. Hodges *et al.*, "Optical investigation of degradation mechanisms in AlGaN/GaN high electron mobility transistors: Generation of non-radiative recombination centers," *Appl. Phys. Lett.*, vol. 100, no. 11, pp. 1–5, 2012.

[11] D. Bisi *et al.*, "Hot-Electron Degradation of AlGaN/GaN High-Electron Mobility Transistors During RF Operation: Correlation With GaN Buffer Design," *IEEE Electron Device Lett.*, vol. 36, no. 10, pp. 1011–1014, Oct. 2015.

[12] Z. Gao, M. F. Romero, A. Redondo-Cubero, M. A. Pampillon, E. San Andres, and F. Calle, "Effects of $Gd_2O_3$ Gate Dielectric on Proton-Irradiated AlGaN/GaN HEMTs," *IEEE Electron Device Lett.*, pp. 1–1, 2017.

[13] Y. Ohno, T. Nakao, S. Kishimoto, K. Maezawa, and T. Mizutani, "Effects of surface passivation on breakdown of AlGaN/GaN high-electron-mobility transistors," *Appl. Phys. Lett.*, vol. 84, no. 12, pp. 2184–2186, Mar. 2004.

[14] Jungwoo Joh and J. A. del Alamo, "Critical Voltage for Electrical Degradation of GaN High-Electron Mobility Transistors," *IEEE Electron Device Lett.*, vol. 29, no. 4, pp. 287–289, Apr. 2008.

[15] R. Jiang *et al.*, "Multiple Defects Cause Degradation After High Field Stress in AlGaN/GaN HEMTs," *IEEE Trans. Device Mater. Reliab.*, vol. 18, no. 3, pp. 364–376, Sep. 2018.

[16] G. Meneghesso *et al.*, "Hot-electron-stress degradation in unpassivated gan/algan/gan HEMTs on Sic," *IEEE Int. Reliab. Phys. Symp. Proc.*, pp. 415–422, 2005.

[17] L. Shi, S. W. Feng, C. S. Guo, and H. Zhu, "Degradation and recovery property of Schottky Barrier height of AlGaN/GaN high electron mobility transistors under reverse AC electrical stress," *ICSICT 2012 - 2012 IEEE 11th Int. Conf. Solid-State Integr. Circuit Technol. Proc.*, pp. 12–14, 2012.

[18] J. Chen *et al.*, "High-Field Stress, Low-Frequency Noise, and Long-Term Reliability of AlGaN/GaN HEMTs," *IEEE Trans. Device Mater. Reliab.*, vol. 16, no. 3, pp. 282–289, Sep. 2016.

[19] D. W. Cardwell *et al.*, "Spatially-resolved spectroscopic measurements of Ec - 0.57 eV traps in AlGaN/GaN high electron mobility transistors," *Appl. Phys. Lett.*, vol. 102, no. 19, pp. 0–4, 2013.

[20] Y. S. Puzyrev *et al.*, "Dehydrogenation of defects and hot-electron degradation in GaN high-electron-mobility transistors," *J. Appl. Phys.*, vol. 109, no. 3, 2011.

[21] A. Armstrong, "Investigation of deep level defects in GaN:C, GaN:Mg and pseudomorphic AlGaN/GaN films," 2006.

[22] G. Verzellesi *et al.*, "Influence of buffer carbon doping on pulse and AC behavior of insulated-gate field-plated power AlGaN/GaN HEMTs," *IEEE Electron Device Lett.*, vol. 35, no. 4, pp. 443–445, 2014.

[23] A. F. Wright, "Substitutional and interstitial carbon in wurtzite GaN," *J. Appl. Phys.*, vol. 92, no. 5, pp. 2575–2585, 2002.

[24] M. J. Uren *et al.*, "'Leaky Dielectric' Model for the Suppression of Dynamic $R_{ON}$ in Carbon-Doped AlGaN/GaN HEMTs," *IEEE Trans. Electron Devices*, vol. 64, no. 7, pp. 2826–2834, Jul. 2017.